\documentclass[prc,showpacs,floatfix,twocolumn]{revtex4-1}
\usepackage{amsmath}
\usepackage{epsfig}
\usepackage{color}
\usepackage{endnotes}
\usepackage{textcomp}
\let\footnote=\endnote

\newcommand{\be}{\begin{equation}}
\newcommand{\ee}{\end{equation}}

\newcommand{\ber}{\begin{eqnarray}}
\newcommand{\eer}{\end{eqnarray}}

\newcommand{\bs}[1]{\ensuremath{\boldsymbol{#1}}}

\begin{document}

\title{Energy resolution with the Lorentz integral transform}

\author{Winfried Leidemann$^{1,2}$
  }

\affiliation{
  $^{1}$Dipartimento di Fisica, Universit\`a di Trento, I-38123 Trento, Italy \\
  $^{2}$Istituto Nazionale di Fisica Nucleare, TIFPA,
  I-38123 Trento, Italy
}

\begin{abstract}
A brief outline of the Lorentz Integral Transform (LIT) method is given.
The method is well established  and allows to treat reactions into the many-body
continuum with bound-state like techniques. The energy resolution that can be achieved 
is studied by means of a simple two-body reaction. From the discussion it will
become clear that the LIT method is an approach with a controlled resolution and that 
there is no principle problem to even resolve narrow resonances in the many-body continuum.
As an example the isoscalar monopole resonance of $^4$He is considered. The importance
of the choice of a proper basis for the expansion of the LIT states is pointed out.
Employing such a basis
a width of 180(70) keV is found for the $^4$He isoscalar monopole resonance 
when using a simple central nucleon-nucleon potential model. 
\end{abstract}

\maketitle

\section{Introduction}

In the last years ab initio approaches have gained an ever increasing importance in the calculation 
of bound-state and reaction observables of nuclear systems with a nucleon number $A \ge 4$ 
(for a comparison of various techniques see~\cite{LeO13,CaD14}).
A particularly challenging aspect in such calculations is the proper treatment of resonances. 
Recently, via the EIHH expansion technique~\cite{BaL00}, the Lorentz integral transform (LIT) method 
\cite{EfL94,EfL07} was applied to calculate  the
isoscalar monopole strength of the $^4$He$(e,e')$ reaction with realistic nuclear forces
paying special attention to the role of the low-lying $0^+$ resonance of $^4$He~\cite{BaB13}. While
the resonance strength could be obtained, it was not possible to determine
the width of the resonance. As it was explained the origin of this deficiency arises from the too
low density of LIT states in the resonance region, 
but it was not really clear why the density happens to be so low.
One may think of various possible reasons: (i) the LIT method is based on bound-state techniques
and might be incapable to describe the shape of narrow resonances, (ii) the used expansion in hyperspherical
harmonics (HH) was not extended to a sufficiently large basis, or (iii) the HH expansion itself is not able
to have a sufficient density of LIT states in the resonance region. The first possibility
can be safely ruled out. In fact in \cite{Le08} it was shown that for a fictitious
two-nucleon system with a narrow resonance in the $^3P_1$ partial wave exact information about
the shape of the resonance could be obtained with the LIT method. Thus the aim of the present paper
is to study which of the two remaining reasons is responsible for the partial failure of the LIT
method in \cite{BaB13} and how one can improve the calculation for having a precise information
on the resonance width.

To clarify the matter first a classical two-body problem, namely deuteron
photodisintegration in unretarded dipole approximation, is discussed. It is shown how the 
density of  LIT states can be increased in this case. In a next step it is considered how one can
achieve such an increase also for the above mentioned isoscalar monopole strength of the $^4$He$(e,e')$ 
reaction. In order to do so a central NN potential model (TN potential) is used that had been previously 
employed in the very first LIT applications for the calculation of the electromagnetic break-up of $^4$He by 
electrons and real photons~\cite{EfL97a,EfL97b}. It is interesting to note that in \cite{EfL97a} one already 
finds a clear signal of the $0^+$ resonance in the $^4$He longitudinal response function, however, at
that early stage of LIT applications the resonance was not studied in greater detail. 

The paper is organized as follows. Section~II contains a brief description of the LIT method.
In section~III the question of the density of the  LIT states 
is addressed and illustrated for the above mentioned two-body case. A LIT calculation for the isoscalar 
monopole strength of the $^4$He$(e,e')$ is discussed in section~IV, where it will be shown how an HH
calculation can be modified in order to determine the width of narrow resonances. Finally, a summary
is given in Section~V.

\section{Lorentz Integral Transform (LIT)}

Over the years the LIT approach~\cite{EfL94} has been applied to a variety of inelastic electroweak reactions.
A rather large number of applications can be found in the review articles \cite{EfL07,LeO13}. 

The LIT of a function $R(E)$ is defined as follows
\begin{equation}
L(\sigma) = \int dE \, {\cal L}(E,\sigma) \, R(E) \,, 
\end{equation}
where the kernel ${\cal L}$ is a Lorentzian,
\begin{equation}
{\cal L}(E,\sigma) = {\frac {1}{(E-\sigma_R)^2 + \sigma_I^2}}
\end{equation}
($\sigma = \sigma_R + i \sigma_I$); the parameter $\sigma_I$
controls the width of the Lorentzian. Because of the adjustable width, and different from 
many other integral transforms, the LIT is a transform with a controlled
resolution. However, aiming at a
higher resolution by reducing $\sigma_I$ might make
it necessary to increase the precision of the calculation. This point will be discussed in greater detail 
in section~III.

For inclusive reactions the LIT $L(\sigma)$  is calculated by solving an equation of the form
\begin{equation} 
\label{eqLIT}
(H-\sigma) \, \tilde\Psi = S \,,
\end{equation}
where $H$ is the Hamiltonian of the system under consideration and $S$ is an
asymptotically vanishing source term related to the operator inducing the  specific reaction. The solution $\tilde\Psi$ is localized.
This a very important property, since it allows to determine $\tilde\Psi$ with bound-state methods,
even for reactions where the many-body continuum is involved. 

Having calculated $\tilde\Psi$ one obtains the LIT from the following expression
\begin{equation}
L(\sigma) = \langle \tilde\Psi | \tilde\Psi \rangle \,.
\end{equation}
The response function $R(E)$ is determined from the calculated $L(\sigma)$ by inverting 
the transform. A general discussion of the inversion and details about various inversion methods
are given in \cite{EfL07,Le08,AnL05,BaE10}.

An alternative way to express the LIT is given by
\begin{equation}
\label{LITeq}
L(\sigma) = - {\frac {1}{\sigma_I} } 
 Im \Big(\langle S | {\frac {1}{\sigma_R + i \sigma_I - H} } | S \rangle \Big) \,.
\end{equation}
This reformulation is useful since it allows a direct application of the Lanczos
algorithm for the determination of $L(\sigma)$ \cite{MaB03}.
In fact the calculations discussed in the following sections are performed 
using expansions on basis sets with a subsequent use of the Lanczos technique.

In order to calculate a specific reaction one has to specify the source term $S$ in Eqs.~(\ref{eqLIT}) and
(\ref{LITeq}). In case of an inclusive electroweak reaction response functions have the general form
\begin{equation}
\label{response}
R(E_f) = \int dE_f |\langle f| \theta | 0 \rangle|^2 \delta(E_f-E_0-\omega) \,,
\end{equation} 
where $| 0 \rangle$ and $| f \rangle$ are ground-state and final-state wave functions of the system under
consideration, while $E_0$ and $E_f$ are the corresponding eigenenergies and $\omega=E_f-E_0$ is
the energy transferred to the system, finally, $\theta$ is the specific 
transition operator that induces the reaction. Note that different from the normal convention here the
argument of the response function $R$ is $E_f$ and not $\omega$.

The source term $S$ for inclusive reactions has the following form
\begin{equation}
| S \rangle = \theta |0 \rangle \,.
\end{equation}
 
A solution of the LIT equation~(\ref{LITeq})
via an expansion on a basis with $N$ basis functions can be understood as follows. One determines the spectrum
of the Hamiltonian for this basis thus finding $N$ eigenstates $\phi_n$ with eigenenergies $E_n$. 
The energies $E_n$ define the positions of the above mentioned LIT  states. Furthermore, the LIT solution
assigns to any eigenenergy a Lorentzian with strength $S_n$ and width $\sigma_I$. It should
be noticed that the source term $|S\rangle$ solely affects the strength leading to 
\begin{equation}
S_n = |\langle \phi_n| \theta | 0 \rangle |^2 \,. 
\end{equation}
The LIT result then reads
\begin{equation}
\label{LIT_En}
 L(\sigma) = \sum_{i=1}^N {\frac {S_n}{(\sigma_R-E_n)^2 + \sigma_I^2}} \,.
\end{equation}
Note that this result is related to the so-called Lanczos response $R_{\rm Lnczs}$ by
\begin{equation}
 R_{\rm Lnczs}(\omega,\sigma_I) = {\frac {\sigma_I}{\pi} } L(\sigma_R=E_0+\omega,\sigma_I) \,. 
\end{equation}
In the limit $\sigma_I \rightarrow 0$ the Lanczos response is equal to the true response function $R$.
However, one often calculates $R_{\rm Lnczs}$ for a small but finite $\sigma_I$ value and identifies
the Lanczos response with the true response, which in general is an uncontrolled approximation.
In the LIT approach one does not make such an identification of transform and response
function. A proper treatment requires an inversion (see discussion in \cite{EfL07}). 

It is important to note that the definition of the LIT in Eq.~(1)
contains the full response function $R$ with all break-up channels. 
For the calculation of the LIT one may use any complete localized $A$-body basis set. 
Automatically for any such set strength from all break-up channels is contained in the LIT.
However, in principle, it can happen that in a specific energy interval of a given reaction
one basis set is more advantageous than another one. 
In fact, such a case is discussed in section~IV.

\section{A simple two-body problem}

To better illustrate the LIT energy resolution 
a simple two-body case is discussed, namely deuteron photodisintegration in unretarded dipole approximation. 
The corresponding cross section is given by
\begin{equation}\label{unret}
 \sigma_{\rm unret}(\omega) = 4 \pi^2 \, \alpha \, \omega \, R_{\rm unret}(E_f=E_0+\omega) \,, 
\end{equation}
where $\omega$ denotes the photon energy and $\alpha$ is the fine structure constant.
The transition operator for the response function $R_{\rm unret}$
is the dipole operator,
\begin{equation} 
\theta = \sum_{i=1}^2 z_i \, {\frac {(1+\tau_{i,z})}{2}} \,,
\end{equation}
 where $z_i$ and $\tau_{i,z}$ are the z-components
of the position vector and of the isospin operator of the ith nucleon, respectively. 
In case of the deuteron the dipole operator induces only transitions to the $np$ final
states $^3P_0,$ $^3P_1$, and $^3P_2$-$^3F_2$. For simplicity in the following example
only transitions to the $^3P_1$ partial wave are considered.
The ansatz for the corresponding $\tilde\Psi$ reads
\begin{equation}\label{3p1_a}
|\tilde\Psi\rangle = R(r) \,\,|(l=1,S=1)j=1\rangle \,|T=1\rangle\,,
\end{equation}
where $r$, $l$, $S$, $j$, and $T=1$ is the relative distance, orbital angular momentum, total spin,
total angular momentum, and isospin of the $np$ pair, respectively.
The resulting LIT equation 
can be easily solved by direct numerical methods or by expansions
of $R(r)$ on a complete set. 
For nuclei with $A>2$ very often 
HH expansions are used with separate hyperspherical and hyperradial parts, where the latter is usually expanded
in Laguerre polynomials $L_n^{(m)}$  times an exponential fall-off.
Therefore a corresponding ansatz is made here for R(r),
\begin{equation}
\label{3p1}
R(r) = r \sum_{n=0}^N c_{n} \,  L_n^{(1)}(r/b) \, \exp\left(-{\frac{r}{2b}}\right) \,,
\end{equation}
where $c_{n}$ are the expansion coefficients and $b$ is a parameter regulating
the spatial extension of the basis. 

\begin{figure}
\centerline{\includegraphics[width=0.48\textwidth]{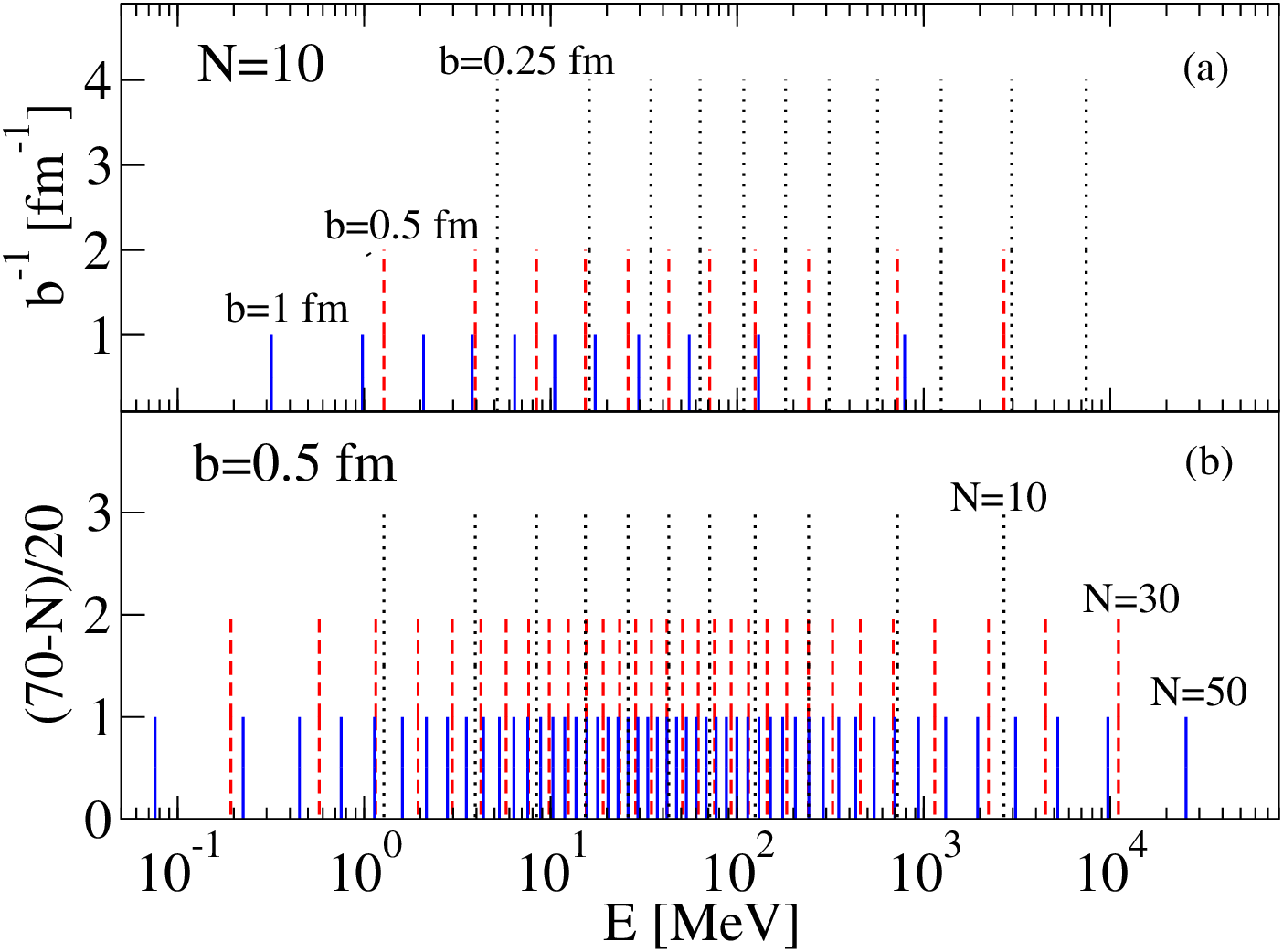}}
\caption{(Color online) Spectrum of the Hamiltonian eigenenergies for the $^3P_1$ NN channel with the AV18 NN
potential for the basis of
Eqs.~(\ref{3p1_a},\ref{3p1}) with various values of $N$ and $b$: $N=10$ with $b=1$ fm, 0.5 fm,
and 0.25 fm in (a) and $b=0.5$ fm with $N=10$, 30, and 50 in (b).} 
\end{figure}
For the following LIT results of the $^3P_1$ channel 
the AV18 NN potential~\cite{AV18} is used. First the Hamiltonian eigenvalues $E_n$,
entering in Eq.~(\ref{LIT_En}), are studied for various basis sets. In the upper panel of Fig.~1 
results are shown with 11 basis functions ($N=10$) and different values for the extension parameter $b$.
One sees that a greater spatial extension of the basis functions leads to a shift of the spectrum
to lower energies. As lowest (highest) eigenenergies one finds 5.17 MeV (7424 MeV), 1.27 MeV (2689 MeV),
and 0.32 MeV (790 MeV) for $b=0.25$, 0.5, and 1 fm, respectively. To obtain a higher density of
LIT states one has to increase the number of the basis states $N$. This is illustrated in the 
lower panel of Fig.~1, where the cases with $N=10$, 30, and 50 are shown for $b=0.5$ fm. 
It is evident that the increase of basis functions does not only lead to a higher density of LIT states, but also
to an extension of the eigenvalues $E_n$ both to lower and higher energies. In fact as lowest (highest)
values one has now 1.27 MeV (2689 MeV), 0.19 MeV (11056 MeV), and 0.076 MeV (25457 MeV) for $N=10$, 30, and 50,
respectively. If one chooses the energy range up to pion threshold one finds that in all three cases
about two-thirds of the $N$ LIT states are located therein.

\begin{figure}
\centerline{\includegraphics[width=0.48\textwidth]{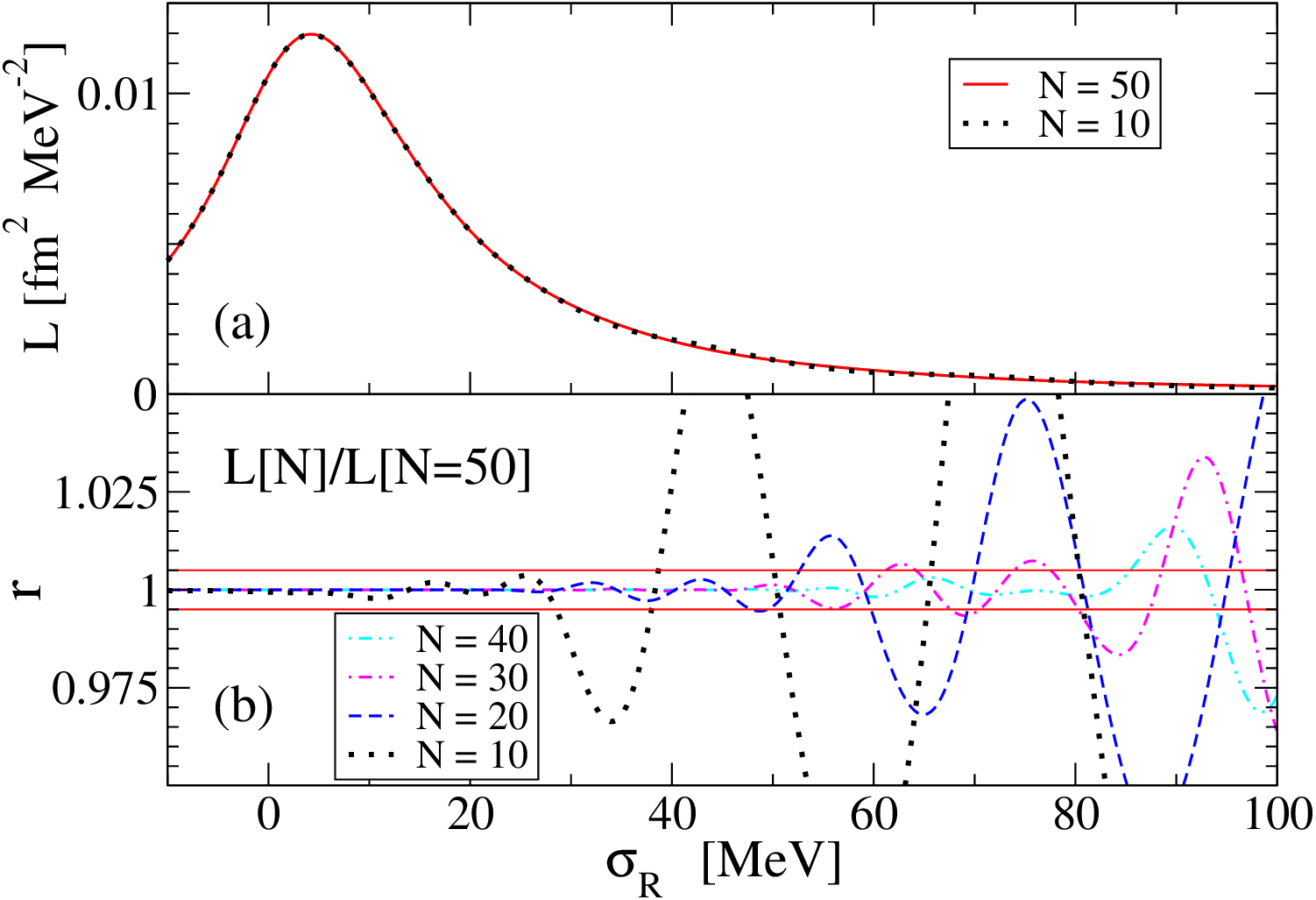}}
\caption{(Color online) LIT $L(\sigma)$ of the $^3P_1$ channel with $\sigma_I=10$ MeV and $b=0.5$ fm. (a): $N=10$ and 50; 
(b): the ratio $r$ of the LIT with $N=10$, 20, 30, and 40 to the LIT with $N=50$ 
(the values 0.995 and 1.005
are illustrated by full lines).} 
\end{figure}
After having discussed the energy distribution of the LIT states we come to the transform $L(\sigma)$ itself.
First, in Fig.~2, the case with $\sigma_I=10$ MeV and $b=0.5$ fm is illustrated. The upper panel of the figure
shows that an increase of the number of basis functions from 11 ($N=10$) to 51 ($N$=50) does
not lead to noticeable differences. In other words, in this case, the LIT is already quite well
converged with a rather small basis. This nice agreement of both results
is not obvious, since Fig.~1 exhibits rather different positions and  densities of the LIT states for both cases.
This has to be interpreted as follows. A discretization of the continuum has no direct physical meaning
and leads in principle to random results. On the contrary one can use the discretization to calculate integral
transforms, as for example the LIT, which lead in convergence to a unique result despite of the randomness of the discrete
eigenenergies.

In order to find differences between the two results of Fig.~2a one has to present them in a different way, as is done
in Fig.~2b. There one sees that the agreement is extremely good up to about 30 MeV and thus one can consider the LIT
for $\sigma_I=10$ MeV to be already converged in this energy range with only very few basis functions.
Beyond 30 MeV the LIT is not yet converged with  $N=10$ and starts oscillating about the LIT with $N=50$ with differences 
becoming even greater than 5\%. Because of this regular oscillations about the more precise result it should even not lead 
to serious inversion problems as long the response does not contain any specific structure at higher energies.
If one wants to check the existence of such structures one should improve the precision of the calculation
by enlarging the basis.
In fact with larger $N$ values of 20, 30, and 40 one finds an increased high-precision range with relative differences 
compared to the case with $N=50$ of less than 0.5\% up to about 50, 60, and 85 MeV, respectively.  
However, in order to search for possible structures it is not the proper strategy to just enlarge the basis.
One has to consider that a resonance with a width considerably smaller than $\sigma_I$ 
is smoothed out in the LIT such that the details of the shape are hidden in tiny contributions to the transform.
To disentangle the details one should reduce $\sigma_I$ increasing in this way the energy resolution of the LIT.
 
The situation for smaller values of $\sigma_I$ is illustrated in Fig.~3, where
LIT results are shown up to about pion threshold for $\sigma_I=0.1$, 1, 2.5, and 10 MeV
and for various values of $N$. For the smallest value, $N=10$ (Fig.~3a), one finds isolated Lorentzian peaks. In case of
$\sigma_I=0.1$ MeV they appear already at low energy, for $\sigma_I=1$ MeV at somewhat higher energy, and
for $\sigma_I=2.5$ MeV at even higher energy. Since the density of LIT states grows with growing $N$
(see other panels of Fig.~3) isolated peaks are pushed to higher and higher energies if $N$ is increased. 
In addition one notes that any decrease 
of $\sigma_I$, $i.e.$ any increase of the resolution, shrinks the convergence range for a given value of 
$N$. For the smallest $\sigma_I$ value of 0.1 MeV a rather strong oscillatory behavior is still present at very low 
energies even with $N=50$. In order to get a smooth and converged LIT also in this case one would need to increase $N$ considerably.

\begin{figure}
\centerline{\includegraphics[width=0.48\textwidth]{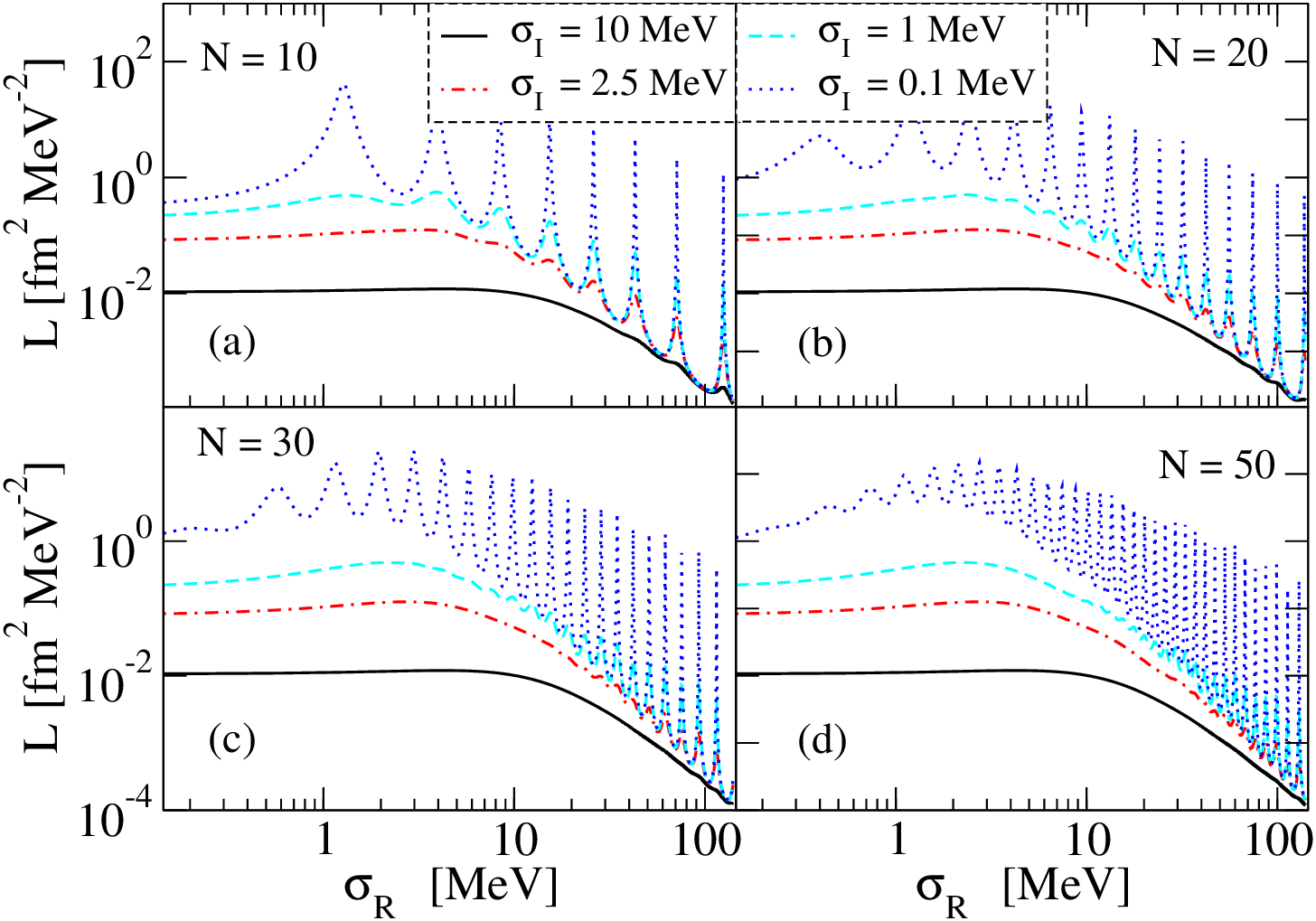}}
\caption{(Color online) LIT $L(\sigma)$ of the $^3P_1$ channel with $\sigma_I=0.1$, 1, 2.5, and 10 MeV in
all four panels ($b=0.5$ fm), but different $N$ values in the various panels: 10 (a), 20 (b), 30 (c), and 50 (d).} 
\end{figure}
For a reliable inversion of $L(\sigma)$ the transform has to be sufficiently converged for a given $\sigma_I$. In 
particular isolated peaks of single Lorentzians should not appear, $i.e.$ for any $\sigma_R$ value one should 
have a significant contribution from more than one Lorentzian. 
Of course, for the present two-body
case one could increase $N$ further without greater problems, but in a general many-body calculation
this might not be easily possible. On the other hand it is not necessary to work with a single
$\sigma_I$ value only. Considering again the LIT results of Fig.~3 with $N=50$ one sees that for $\sigma_I=1$ MeV
a rather converged result is obtained up to about 10 MeV. Thus structures of
a relatively small width would leave a visible signal in the transform in this energy range.
In fact one can use this LIT in the low-energy range and combine it with a LIT with a larger $\sigma_I$ for higher energies.
In general one can proceed as follows, one defines a new
transform $T$ in an interval $[\sigma_0,\sigma_M]$ by combining $M$ transforms $L_m=L(\sigma_R,\sigma_{I,m})$ such that only the 
LIT with the specific  $\sigma_I=\sigma_{I,m}$ enters significantly in the energy range $[\sigma_{R,m-1},\sigma_{R,m}]$: 
\begin{equation}\label{LIT_mult}
T = L_1 f_1 
  + \sum_{m=2}^{M-1} a_m L_m (1-f_{m-1})) f_m 
+ a_M L_M (1-f_{M-1}) \,.
\end{equation}
The function $f_m$ in Eq.~(\ref{LIT_mult}) is a smooth cutoff from 1 to 0 at $\sigma_R = \sigma_{R,m}$,
for example $f_m(\sigma_R \le \sigma_{R,m})=1$ and $f_m(\sigma_R > \sigma_{R,m})= \exp(-((\sigma_R-\sigma_{R,m})/\Delta)^{2n})$,
where one has to take reasonable values for the parameters $\Delta$ and $n$. The coefficients $a_m$ in Eq.~(\ref{LIT_mult})
can be chosen such that size of the transform does not change too drastically from one energy range to the next.
Thus, for the case of the LITs of Fig.~3 with $N=50$, one could set $\sigma_{I,1}=1$, $\sigma_{I,2}=2.5$, and 
$\sigma_{I,3}=10$ MeV with $\sigma_{R,1}=10$ and $\sigma_{R,2}=30$ MeV. In order to improve the precision
in the threshold region even further one could include the case with 
$\sigma_I=0.1$ MeV, but one would need to further increase $N$.
Alternatively, keeping $N=50$,
one could check the convergence behavior of a LIT with a somewhat larger $\sigma_I$, for example $\sigma_I=0.25$
MeV.

The discussion above shows that the LIT is an approach with a controlled resolution.
In an actual calculation one should check which is the lowest $\sigma_I$ that leads in a specific
energy range to a sufficiently converged and smooth LIT without that a single LIT state sticks out. 
Structures which are considerably smaller than such a $\sigma_I$ value cannot be resolved by the inversion. 
A helpful criterion is given in \cite{Le08} (see discussion of Fig.~7 in \cite{Le08}).

\section{$^4$He Isoscalar monopole response function}  

The isoscalar monopole response function $M(q,E_f=E_0+\omega)$ can be determined in inclusive
electron scattering ($q$ and $\omega$ represent momentum and energy transferred by
the virtual photon to the nucleus). In this case the transition operator $\theta$ of Eq.~(\ref{response}) 
becomes $q$-dependent and takes the form
\begin{equation}\label{monopole}
 \theta(q) =\frac{G_E^s(q^2)}{2} \sum_{i=1}^A \, j_0(q r_i)\,,
\end{equation}
where $G_E^s(q^2)$ is the nucleon isoscalar electric form factor, 
$\bs{r}_i$ is the position of nucleon $i$,
and  $j_0$ is the spherical Bessel function of 0$^{th}$ order.

Experimental investigations of the $^4$He$(e,e')$ reaction in \cite{Wa70,Fr65,Ko83} revealed a 
$0^+$ resonance located less than 1 MeV above the $^4$He break-up threshold with quite a narrow 
width of about 250 keV.  It is interesting to note that
very recently a EIHH-LIT calculation was carried out, where it is pointed out that the resonance might be 
interpreted as a breathing mode \cite{BaB15}. Unfortunately, this and the preceding EIHH-LIT calculations~\cite{BaB13}
were only able to determine a resonance strength but not a resonance width  
since the density of LIT states in the resonance region was too low.
On the other hand, with the experience made in the previous section it seems to be easy to 
increase the density of LIT states also in the region of the $^4$He isoscalar monopole resonance, namely 
by increasing the number of HH basis states. Unfortunately with an HH expansion this does not work very well
in the energy region below the three-body break-up threshold. To illustrate this, 
the first LIT application~\cite{EfL97a}, already mentioned in the introduction, is considered in the following.
The calculation was carried out for inclusive electron scattering off $^4$He and used a correlated HH (CHH) basis,
where NN short range correlations are introduced to accelerate the convergence of the HH expansion. 
The NN interaction (TN potential, Coulomb force included) consisted in a central spin-dependent 
interaction active only in even NN partial waves. In Fig.~4 we show an unpublished
result for the LIT of the response function $M$ from this calculation.
One sees that there is only an isolated LIT state in the resonance region. In this specific case
one finds the next LIT state only about 0.1 MeV below the four-body break-up threshold at $\sigma_R=0$ MeV. On the contrary for
positive values of $\sigma_R$ one has quite a high density of LIT states. An increase of the number
of hyperspherical and/or hyperradial basis states does not change the general picture of having
only very few LIT states for negative $\sigma_R$ values, as in fact it was the case in the LIT-EIHH calculations~\cite{BaB13,BaB15}. 
In addition it was checked for the present work that an increase of the parameter $b$ in the hyperradial 
functions ($\sim L_n^{(8)}(\rho/b) \, \exp(-\rho/2b))$ in the four-body CHH calculation of \cite{EfL97a}
does not lead to any significant change concerning the low-energy density of LIT states.
\begin{figure}
\centerline{\includegraphics[width=0.4\textwidth]{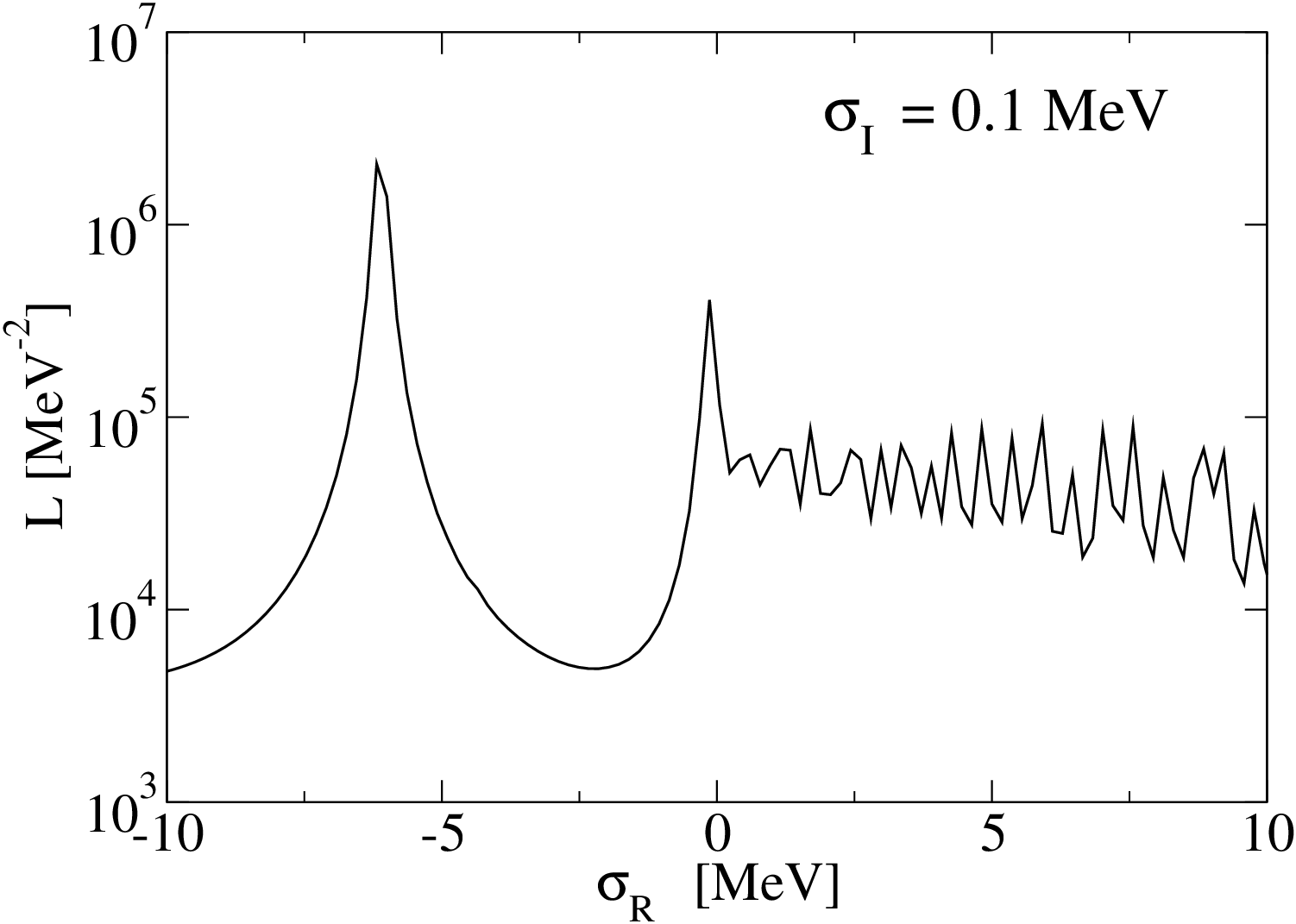}}
\caption{LIT $L(\sigma)$ of $M(q,\omega)$ at $q=300$ MeV/c (TN potential) of $^4$He with four-body CHH basis from~\cite{EfL97a}.} 
\end{figure}

At first sight it is not understandable why an increase of the HH basis states does not have
a significant effect on the density of LIT states in the energy range below the three-body break-up
threshold. On the other hand one has to consider that the dynamical variable for a two-body break-up, 
$i.e.$ the relative vector ${\bf r}^\prime_4 = {\bf r}_4 - \sum_{i=1}^3 {\bf r}_i/3$ between the free nucleon and the bound
three-body system, does not appear explicitly
in the HH formalism. Therefore it might be helpful to use a different basis, where ${\bf r}^\prime_4$ 
is taken into account as variable explicitly. In the following, the four-body calculation is switched to such a basis.
The new basis states consist in a product of a three-body CHH basis state for the first three nucleons 
times a single-particle basis state for the fourth nucleon. The CHH basis state is then again a product of a hyperspherical basis state
${\cal Y}_{[K\alpha]}(\Omega)$ times a hyperradial basis state $R_n(\rho)$, where the former
depends on the grand-angle $\Omega$ and is characterized by the grand-angular quantum number $K$ and 
a set of other quantum numbers $[\alpha]$, whereas the latter is proportional to $L_n^{(5)}(\rho/b) \, exp(-\rho/2b)$
(for a detailed definition of the HH basis see for example \cite{BaL00}).

The single-particle states are defined as follows 
\begin{equation} \label{basis}
|\Phi({\bf r}^\prime_4)\rangle = \sum_{l_4 m_{s_4} m_{t_4}} \phi_{l_4}(r^\prime_4) \,|l_4 m_{l_4}\rangle \, 
 |s_4 m_{s_4}\rangle\,
 |t_4 m_{t_4}\rangle\,,
\end{equation}
where $l_4$, $s_4=1/2$ and $t_4=1/2$ are the orbital angular momentum, spin, and isospin quantum numbers, respectively,
and $m_{l_4}$, $m_{s_4}$ and $m_{t_4}$ denote the corresponding projections. The radial wave function $\phi_{l_4}(r^\prime_4)$
is given by
\begin{equation}\label{rel_r}
\phi_{l_4}(r^\prime_4) = (r^\prime_4)^{l_4} \prod_{i=1}^3 f(r_{i4})  \sum_{n_4=0}^{N_4-1} c_{n_4} 
L_{n_4}^{(2)}(r^\prime_4/b_4) \exp\left({\frac{-r^\prime_4}{2b_4}}\right),
\end{equation}
where $f(r_{i4}) = f(|{\bf r}_i - {\bf r}_4|)$ is the NN correlation function (same correlation as in CHH basis).
The CHH basis provides antisymmetric
states for the first three nucleons, whereas the product of the CHH basis states and the single-particle states
has to be antisymmetrized in order to have totally antisymmetric basis states for all four particles.
In order to calculate the $^4$He ground state or the transitions induced by the action of the operator $\theta(q)$ of 
Eq.~(\ref{monopole}) on the $^4$He ground state one has to couple the CHH basis states with those of Eq.~(\ref{basis}) 
to a total angular momentum $L$ equal to zero, and also the total spin (isospin) wave function of the four nucleons
has to be coupled to a total spin $S$ (isospin $T$) equal to zero. 

The calculations described in the following proceed in the same way as it was carried out in \cite{EfL97a,EfL97b},
namely by using a nine-dimensional Monte Carlo integration for the evaluation of the various Hamiltonian
and norm matrix elements. Such an approach leads to reliable results (see benchmark test~\cite{BaF01}). 

\begin{table}
\caption{Convergence of $^4$He binding energy BE with the TN potential (Coulomb force included): $l_4$ denotes the orbital angular momentum of
the single-particle motion (see Eqs.~(\ref{basis},\ref{rel_r})), $K_3$ is the grand-angular quantum number of the three-body CHH states,
``sym'' and ``mixed'' indicate that the CHH state is symmetric and mixed symmetric, respectively 
(note that for $l_4>0$ two symmetric CHH states
and only the lowest mixed symmetric state are taken).}\label{tab:BE1}
\begin{center}
\begin{tabular}{c|c|c|c|c} \hline\hline
$l_4$ & $K_3$ \,\,\, & \,\,\,CHH symmetry \,\,\, &  \# CHH states & BE [MeV]   \\
\hline
  0 &  0  &  sym  & 1 &  28.66  \\
  0 &  4  &  sym  & 1 &  30.11  \\
  0 &  6  &  sym  & 1 &  30.56  \\
  0 &  8  &  sym  & 1 &  30.67  \\
  0 &  10  &  sym  & 1 &  30.77  \\
  0 &  12  &  sym  & 2 &  30.82  \\  
  0 &  14  &  sym  & 1 &  30.85  \\
  0 &  2  &  mixed      & 1 &  31.28  \\
  0 &  4  &  mixed      & 1 &  31.32 \\
  1 & 1,3,5 & sym and mixed  & 3 & 31.39 \\
  2 & 2,4 & sym and mixed  & 3 & 31.41 \\

\hline\hline
\end{tabular}
\end{center}
\end{table}
In Table~\ref{tab:BE1} the convergence of the $^4$He binding energy is illustrated.
It is evident that the dominant contribution comes from the basis functions where the three-body CHH basis
is in a symmetric state and the relative angular momentum $l_4$ of the single-particle motion is equal to zero. 
Mixed symmetric CHH states together with $l_4=0$ enhance the binding energy by about 0.5 MeV. The contribution
of states with $l_4>0$ is very small and leads to a further increase of about 0.1 MeV
(antisymmetric CHH states have been neglected).
The final result of 31.41(5) MeV agrees very well with the converged result of the four-body CHH calculation of 31.40(5) MeV.

For the calculation of the LIT of the $^4$He isoscalar monopole response function $M(q,\omega)$
hyperradial and radial basis states are chosen with a rather large extension in space. This should lead to a sufficiently
high density of LIT states at low energy (see Fig.~1). The following choice is made: $b=4$ fm (hyperradial CHH states) 
and $b_4=1.33$ fm (radial single-particle states). 
For the hyperradial part 15 basis functions are taken, whereas the number of the radial single-particle basis states is 
kept variable and denoted by $N_4$.
Since the main interest of this investigation is concentrated on the low-energy part of $M(q,\omega)$  only
those three-body CHH basis states which lead to a significant contribution to the 
$^4$He binding energy (see Table~\ref{tab:BE1}) are taken into account: symmetric CHH states up to $K_3 = 6$ and the mixed symmetric CHH state 
with $K_3 = 2$. All the remaining states contribute only with 0.42 MeV to the $^4$He binding energy. Thus
it is reasonable to expect also a shift of the position of the $^4$He $0^+$ resonance by about 0.5 MeV towards higher
energies with respect to the four-body CHH calculation of Fig.~4 leading to a value of about -5.9 MeV.
To take into account only four hyperspherical three-body CHH states reduces the numerical effort of the 
calculation quite a bit, but even with such a reduced
basis the calculation requests a considerable numerical effort (note that due to the rather weak fall-off of the hyperradial/radial
basis functions the relevant integration volume becomes much larger than for the bound-state calculation).

\begin{figure}
\centerline{\includegraphics[width=0.4\textwidth]{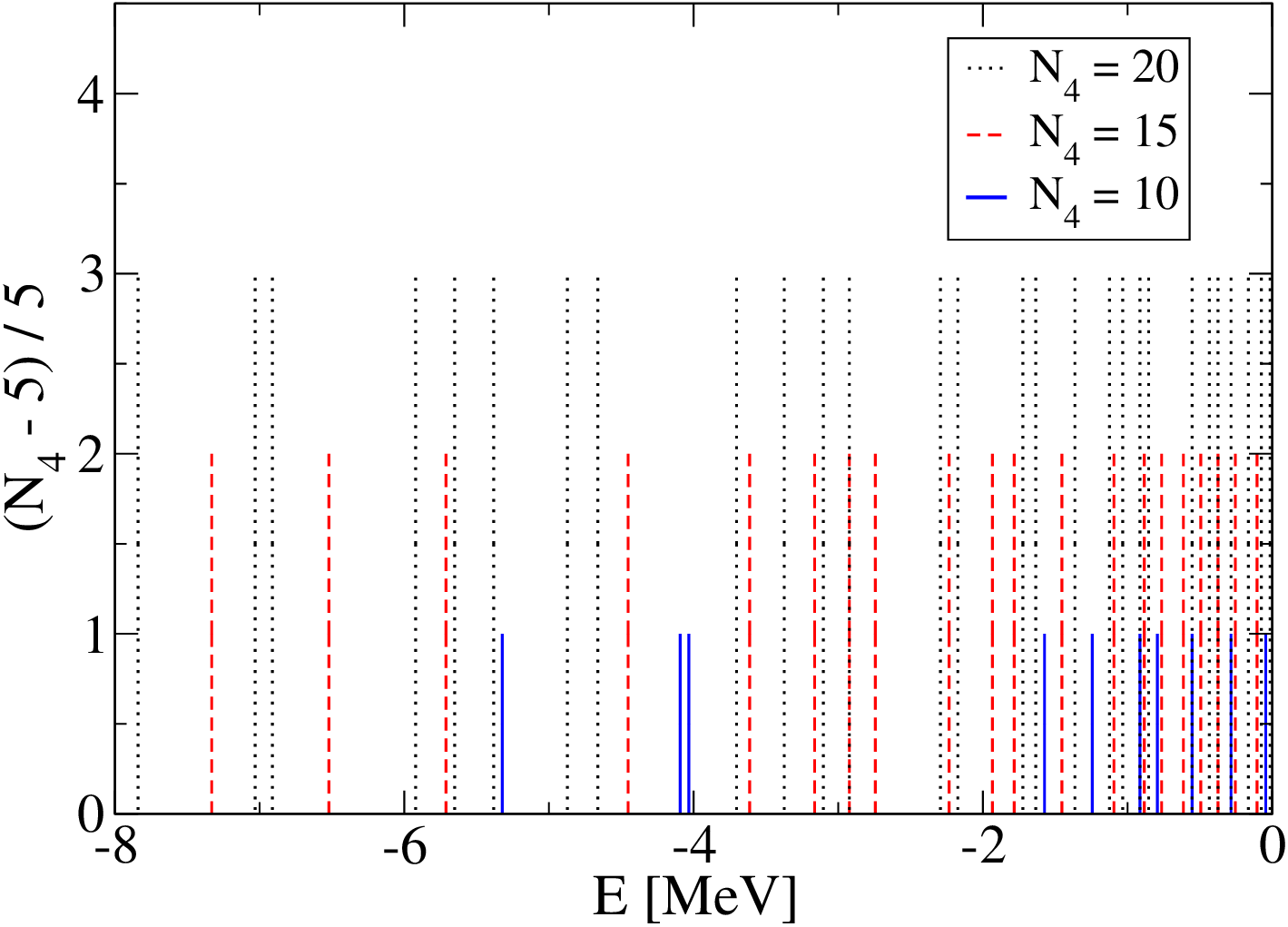}}
\caption{(Color online) Spectrum of the Hamiltonian eigenenergies for the $^4$He$(J^\pi=0^+)$ states with the TN NN
potential for the basis described in the text with $N_4$ basis functions for the radial single-particle basis.} 
\end{figure}

In Fig.~5 the energy distribution of LIT states is shown up to the four-body break-up threshold
with three different $N_4$ values. As it has been anticipated
the density of LIT states increases with a growing number of radial single-particle states. 
Due to the interplay of CHH basis states and radial single-particle
basis states the pattern is not as regular as in the two-body example
of Fig~1. 

It is obvious that a further increase of $N_4$ would lead to an even higher density of low-energy 
LIT states, but, unfortunately, the precision of the nine-dimensional Monte Carlo integration is not sufficiently high and
a solution with $N_4>20$ becomes problematic. In this respect it is a drawback to work with a correlated
HH basis because one looses the orthogonality. Nonetheless, as shown in the following,
the present calculation allows a determination of the width of the $^4$He isoscalar monopole resonance.

In Fig.~6 the LIT of $M(q,\omega)$ is shown for $N_4=20$ and $N_4=22$. In the latter case
the basis function with $n_4=20$ (see Eq.~(\ref{rel_r})) is dropped because it makes the numerical solution
of the corresponding eigenvalue problem problematic (precision of Hamiltonian and norm matrices is not sufficiently high). 
The figure illustrates that with increasing resolution,
$i.e.$ with decreasing $\sigma_I$ the resonance becomes more and more pronounced against the background. One also finds
that the anticipated peak position of about -5.9 MeV is roughly confirmed. 

In order to determine a resonance width the LIT has to be inverted. The procedure of the
inversion in presence of a resonance is described in \cite{Le08}.
As rule of thumb one can say that the chosen $\sigma_I$ should not be much larger than the resonance width. However, in principle,
if the LIT is calculated with a very high precision one can choose also a considerably larger $\sigma_I$. In fact in the model study
\cite{Le08} it is nicely demonstrated that the obtained width is independent from the used $\sigma_I$ over a very large range,
even with $\sigma_I=5$ MeV the shape of a resonance with a width of 270 keV was exactly determined. In the present 
case the precision of the calculated LIT is not as high as in \cite{Le08}. The actual inversion was made with two different
$\sigma_I$ values (see discussion of transform $T$ introduced in Eq.~(\ref{LIT_mult})), 
$\sigma_{I,2}=5$ MeV beyond -4 MeV, whereas $\sigma_{I,1}$ was varied in the range
from 0.1 to 0.5 MeV in the low-energy region. The various results for the width were quite stable (maximal difference: 0.01 MeV) and lead to the
following values: 120(10) keV ($N_4=20$) and
240 keV ($N_4=22)$, the mean value amounts to 180(70) keV. The result lies in the same ballpark
as the experimental value of 270(50) keV \cite{Wa70}. One could try to increase the theoretical precision of the
determination of the width by a further increase of $N_4$. Because of the problem just mentioned above this would require
a non negligible effort in the present calculation. On the other hand it is much more desirable to make
such a calculation with a realistic nuclear force instead of having a very precise result for a simple NN interaction like
the TN potential model.

For the inversion of the LIT it is assumed that there is only a single peak in the resonance region.
In principle from an increase of the precision of the calculation one could also find out that the resonance has a more 
complicated structure, for example a double peak. It is evident from Fig.~6 that in the present calculation one controls 
the resonance with a resolution of about 100 keV, thus it is not possible to resolve structures with an even smaller width.
The situation is similar to experiment, where one works with a given resolution of the experimental
apparatus.

\begin{figure}
\centerline{\includegraphics[width=0.45\textwidth]{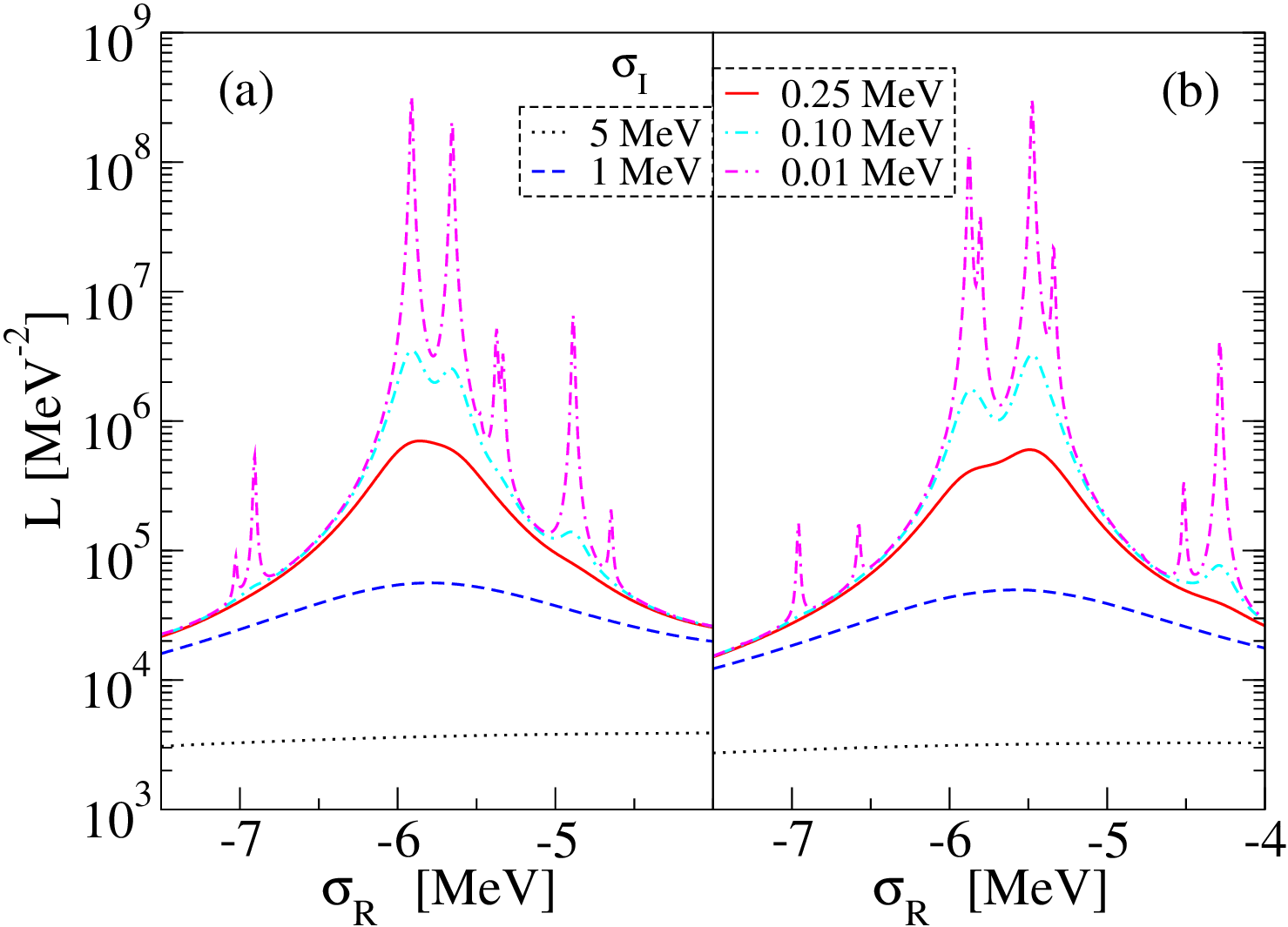}}
\caption{(Color online) LIT $L(\sigma)$ of $M(q,\omega)$ at $q=250$ MeV/c (TN potential) of $^4$He with $\sigma_I=0.01$, 0.1, 0.25, 1, and 5
MeV calculated with the new basis states (three-body CHH basis times
single-particle basis of Eqs.~(\ref{basis},\ref{rel_r})); in (a) the single-particle basis with 20 and in (b) with 21 states (see text).} 
\end{figure}

\section{Summary}

In this work first a brief outline of the LIT approach is given, a method that allows to calculate continuum observables 
with bound-state techniques. In a next step the energy resolution that can be obtained with the method is discussed. 
To this end a specific channel of the final state in deuteron photodisintegration is studied
in unretarded dipole approximation. In order to do so the LIT equation is solved using
a bound-state basis to calculate the relevant quantities. It is shown how the spatial extension
and the number of basis states affects the calculation. The role of the width parameter $\sigma_I$
of the LIT is discussed, especially which $\sigma_I$ values should be chosen
in a specific situation. The discussion makes clear that the LIT constitutes an
approach with a controlled resolution. In particular it allows to determine the width
of narrow resonances in the continuum. However, for LIT calculation with an arbitrary many-body bound-state basis
this is not always guaranteed. Usually narrow resonances of an $A$-body system in the continuum
are located below the three-body break-up threshold.
If one intends to study the resonance in a scattering state calculation it is mandatory
to take into account the relevant {\it dynamical} variable,
$i.e.$ the relative vector between the two fragments.
In spite of the LIT bound-state character this {\it dynamical} variable should also appear explicitly in a LIT calculation. 
It guarantees that the density of LIT states can be enhanced in the resonance region by increasing the number of 
those basis states that directly depend on the {\it dynamical} variable.
On the contrary, if this variable is not included explicitly, as for example in an $A$-body HH calculation, 
it is difficult, maybe even impossible, to obtain a detailed information about the resonance.

In a general case of a resonance with various open channels the resonance can have
a {\it partial} decay width to all these channels, which then results in a {\it total} decay width.
In this case it is sufficient to choose just one of the various possible {\it dynamical} variables.
The only aim which has to be fulfilled is a sufficient density of LIT states.
The LIT then by definition collects strength from all open channels and a determination of the
width via inversion leads to the {\it total} width.

In order to illustrate the situation in greater detail the isoscalar monopole response function $M$ of $^4$He is considered
as test case using a simple spin-dependent central NN interaction (Coulomb force is included). 
In fact, here one finds a narrow resonance in the continuum below the three-body break-up threshold.
Obviously, in this case the {\it dynamical} variable is given by the relative vector of the free nucleon
and the bound three-nucleon system. 
Using a proper basis as described above it is indeed found that the density of LIT states
grows in the resonance region if the number of single-particle basis states is enhanced. 
It is shown that the LIT state density becomes sufficiently high to determinate the resonance width
via an inversion of the transform.
In fact, a width of 180(70) keV is found, a result that is not too far from the experimental
value of 270(50) keV. It would be very interesting to perform such a calculation also
with modern realistic nuclear forces as have been used in the LIT-EIHH calculations of the response function $M$\cite{BaB13,BaB15},
however the effective interaction approach has to be a bit redesigned, since one would not have any more a pure HH basis.  
 
The present approach is not only advantageous in case of
cross sections with narrow resonances, but also for non-resonant two-body break-up cross sections at very low energies.
The possibility to increase the LIT state density allows to work with smaller $\sigma_I$ values thus
enhancing the energy resolution of the calculation. This might be particularly interesting in case of
astrophysical reactions.


\end{document}